\documentclass[conference]{IEEEtran}
\usepackage[T1]{fontenc}
\usepackage[noadjust]{cite}
\usepackage{graphicx,color}
\usepackage[dvipsnames]{xcolor}
\usepackage{amsmath,amsbsy,amsfonts,amssymb,amsthm}
\usepackage{mathrsfs,bm,bbm}
\usepackage{mathtools}
\usepackage{braket}
\usepackage{lipsum}
\usepackage{pgfplots}
\usepackage{pgfplotstable}
\mathtoolsset{showonlyrefs}
\ifCLASSOPTIONcompsoc
\usepackage[caption=false,font=normalsize,labelfont=sf,textfont=sf]{subfig}
\else
\usepackage[caption=false,font=footnotesize]{subfig}
\fi

\IEEEoverridecommandlockouts 

\usepackage{etoolbox}
\usepackage{tikz}
\newrobustcmd*{\squareA}[1]{\tikz{\filldraw[draw=#1,fill=#1] (0,-0)
rectangle (0.1cm,0.14cm);}}
\newrobustcmd*{\mycircle}[1]{\tikz{\filldraw[draw=#1,fill=#1] (0,-0.3) circle [radius=0.08cm];}}

\usepackage{tikz}
\usetikzlibrary{shapes}
\usepackage{braket}
\usepackage{acro}

\usepackage[nopostdot,nogroupskip,style=super,nonumberlist,acronym]{glossaries}

\newacronym{bs}{BS}{Base Station}
\newacronym{qs}{QS}{Quantum Station}
\newacronym{cbs}{CBS}{Classical Base Station}
\newacronym{qbs}{QBS}{Quantum Base Station}
\newacronym{ue}{UE}{User Equipment}
\newacronym{cue}{CUE}{Classical User Equipment}
\newacronym{que}{QUE}{Quantum User Equipment}
\newacronym{qkd}{QKD}{Quantum Key Distribution}
\newacronym{qlan}{QLAN}{Quantum Local Area Network}
\newacronym{qsdc}{QSDC}{Quantum Secure Direct Communication}
\newacronym{prb}{PRB}{Physical Resource Block}
\newacronym{los}{LOS}{Line-of-Sight}
\newacronym{fso}{FSO}{Free Space Optic}

\usepackage{verbatim}

\newcommand{\pp}[1]{\textcolor{black}{#1}} 

\begin{document}

\title{1Q: First-Generation Wireless Systems Integrating Classical and Quantum Communication}

\author{
\IEEEauthorblockN{Petar Popovski,~\IEEEmembership{Fellow,~IEEE}, \v Cedomir Stefanovi\' c,~\IEEEmembership{Senior Member,~IEEE}, Beatriz Soret,~\IEEEmembership{Senior Member,~IEEE},\\ Israel Leyva-Mayorga,~\IEEEmembership{Member,~IEEE}, Shashi Raj Pandey,~\IEEEmembership{Member,~IEEE}, René B. Christensen,\\ Jakob K. S\o ndergaard, Kristian S. Jensen,~\IEEEmembership{Graduate Student Member,~IEEE}, Thomas G. Pedersen,\\ Angela Sara Cacciapuoti,~\IEEEmembership{Senior Member,~IEEE}, and Lajos Hanzo,~\IEEEmembership{Life Fellow,~IEEE}
\thanks{Petar Popovski (petarp@es.aau.dk), \v Cedomir Stefanovi\' c (cs@es.aau.dk), Israel Leyva-Mayorga (ilm@es.aau.dk), Shashi Raj Pandey (srp@es.aau.dk), Jakob K. S\o ndergaard (jakobks@es.aau.dk), and Kristian S. Jensen (ksjen@es.aau.dk) are with the Department of Electronic Systems, Aalborg University, Denmark.}
\thanks{Beatriz Soret (bsoret@ic.uma.es) is with Telecommunication Research Institute, University of Málaga, Spain, and Department of Electronic Systems, Aalborg University, Denmark}
\thanks{René B. Christensen (rene@math.aau.dk) is with the Department of Mathematical Sciences, Aalborg University, Denmark.}
\thanks{Thomas G. Pedersen (tgp@mp.aau.dk) is with the Department of Materials and Production, Aalborg University, Denmark.}
\thanks{Angela Sara Cacciapuoti (angelasara.cacciapuoti@unina.it) is with Quantum Internet Research Group, University of Naples Federico II, Italy.}
\thanks{Lajos Hanzo (hanzo@soton.ac.uk) is with the School of Electronics and Computer Science, University of Southampton, U.K.}
\thanks{This work was supported, in part, by the Danish National Research Foundation (DNRF), through the Center CLASSIQUE, grant nr. 187. Angela Sara Cacciapuoti's work has been funded by the European Union under the ERC grant QNattyNet, n.101169850. Views and opinions expressed are however those of the author(s) only and do not necessarily reflect those of the European Union or the European Research Council. Neither the European Union nor the granting authority can be held responsible for them. L. Hanzo would like to acknowledge the financial support of the Engineering and Physical Sciences Research Council (EPSRC) projects Platform for Driving Ultimate Connectivity (TITAN) under Grant EP/X04047X/1; Grant EP/Y037243/1; and Robust and Reliable Quantum Computing (RoaRQ, EP/W032635/1).}
\thanks{The authors would like to thank Jane Rygaard Pedersen, Nokia, for the initial discussion that led to the term 1Q.}\thanks{This version of the accepted manuscript is licensed under CC BY.}}
}
\maketitle

\begin{abstract}
We \pp{introduce} the concept of 1Q, the first wireless generation of integrated classical and quantum communication. \pp{1Q} features quantum base stations (QBSs) that support entanglement distribution via free-space optical links alongside traditional radio communications. Key new components include quantum cells, quantum user equipment (QUEs), and hybrid resource allocation spanning classical time-frequency and quantum entanglement domains.
Several application scenarios are discussed and illustrated through system design requirements for quantum key distribution, blind quantum computing, and distributed quantum sensing. A range of unique quantum constraints are identified, including decoherence timing, fidelity requirements, and the interplay between quantum and classical error probabilities. Protocol adaptations extend cellular connection management to incorporate entanglement generation, distribution, and handover procedures, expanding the Quantum Internet to the cellular wireless.
\end{abstract}

\section{Introduction}

Wireless mobile communication has evolved through a symbiosis of  \emph{(i)} New services, \emph{(ii)} Specific wireless technologies like MIMO, and \emph{(iii)} Architectural principles. The result is ubiquitous access to the services of the global internet, used for transfer of classical information. The \emph{Quantum Internet (QI)}~\cite{kimble2008quantum} is envisioned as a network capable of processing and transmitting quantum information and it is interesting to assess the three symbiotic elements above in that context. Regarding \emph{(i)}, several generic services have already been identified:  quantum-safe security, distributed quantum computing, and networked quantum sensing. For \emph{(ii)}, many specific quantum technologies, like quantum memories or quantum repeaters, are not mature enough for widespread deployment. Yet, quantum information processing has reached certain maturity - its own \emph{transistor moment}: the transistor was invented in $1947$, but the chips only reached practical scale in the $1960$s, when a transistor became a functional block of a circuit-diagram. In analogy to this, along with the vibrant development of quantum technology, it is time to integrate its functional blocks into larger systems and networks. 

This brings us to \emph{(iii)}, the architectural principles for building a mobile wireless communication network, capable of supporting distributed quantum applications. Our conjecture is that these principles will closely conform to the present mobile wireless infrastructure; yet, with certain revolutionary changes due to the physical nature of the QI. This is reinforced by the fact that networked quantum applications critically hinge on the exchange of classical information, in addition to quantum information~\cite{cacciapuoti2020entanglement}.  

This paper presents the design options and the associated challenges to be addressed as the cellular infrastructure becomes part of the QI. We refer to this system as \emph{1Q}, the first wireless generation of integrated classical and quantum communication. In order to formulate to 1Q's architectural principles, we note that the architecture of wireless mobile communication systems started to develop from a simply formulated objective: \emph{Construct a system that can offer untethered connection to a device moving in space} to dispense with Graham Bell's tethered phones, constraining the movement of users. The foundational concept of modern mobile communication relies on a system of cells illuminated by RF signals from a Base Station (BS). The BS offers wireless connection to devices roaming within its cell as well as negotiates with the other BSs how to hand over users that move from one cell to another.

The 2G-6G evolution has been solely dedicated to the transmission of classical information bits over the wireless medium. 2G supported ubiquitous digital global voice services, complemented by the unlikely hero SMS, launching the texting era. 3G gave birth to the wireless Internet data supported by the smartphone. 4G perfected global connectivity that fuelled social media and flawless video. 5G was conceptualized as a ``connectivity supermarket'', covering diverse connectivity types in terms of rate, latency, or traffic-density. 6G is poised to bring about AI-aided mobile communication, upgrading the task of \gls{bs} from pure communication to gathering sensing and positioning data. Crucially, 6G integrates the terrestrial and non-terrestrial networks. 
Hence, any future mobile network will have integrated support from satellites, critically important for entanglement distribution. 

We \emph{hypothesize that beyond-6G systems will include quantum applications and therefore can be termed as 1Q rather than 7G.} Quantum has two important aspects:
\begin{enumerate}
    \item Classical information transmitted over the wireless system is used to support certain quantum protocols, such as \gls{qkd} or quantum teleportation.
    \item Wireless transmission of quantum bits (qubits). Some of these qubits carry quantum information, while others are used in support of entanglement distribution.
\end{enumerate}
The first aspect can be viewed as a new set of quantum applications that impose unique requirements on classical wireless systems; this positions the 1Q concept as an evolutionary stage of cellular wireless connectivity. Furthermore, QKD may provide quantum-safe security for classical applications. The second aspect is revolutionary, as it is incompatible with the classical wireless communication theory and practices, since resource allocation and management policies are fundamentally changed. 

\pp{The following conceptual exposition sacrifices mathematical rigor in favor of plausible quantum information processing and transmission aspects. The next section introduces the basic principles of classical and quantum wireless links, outlining their differences. Section~\ref{sec:1Qsystemmodel} describes the key architectural elements, classical/quantum Base Stations and cells. Section~\ref{sec:ApplicationScenarios} explores generic quantum application scenarios in the context of 1Q, such as quantum key distribution, quantum computation, and quantum sensing. These are further exemplified in Section~\ref{sec:QKD_examples}, providing detailed examples and system-level requirements, while Section~\ref{sec:architecture} discusses 1Q from an architectural and protocol perspective. Section~\ref{sec:conclusion} concludes the paper.}

\section{Wireless Links: Classical and Quantum}
\label{sec:WirelessLinks}

\begin{figure*}[t!]
 \centering
 \begin{minipage}[b]{0.6\textwidth}
   \includegraphics[page=4]{TransmissionTypes.pdf}\\[8mm]
   \subfloat[]{
       \includegraphics[page=1]{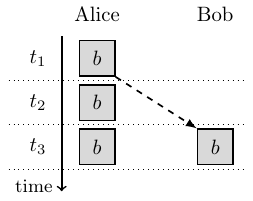}
       \label{fig:transmissionClassic}
   }
   \hspace{1cm}
   \subfloat[]{
       \includegraphics[page=2]{TransmissionTypes.pdf}
       \label{fig:transmissionQuantum}
   }
 \end{minipage}
 \begin{minipage}[b]{0.28\textwidth}
   \subfloat[]{
       \includegraphics[page=3]{TransmissionTypes.pdf}
       \label{fig:transmissionTeleport}
   }
 \end{minipage}
 \caption{Comparison between wireless transmissions in the classical and quantum domains. Each subfigure shows the memories of two users, Alice and Bob, at each timeslot $t_i$. \protect\subref{fig:transmissionClassic}~shows the wireless transmission of a classical bit. \protect\subref{fig:transmissionQuantum}~shows the direct transmission of quantum information using a flying qubit. \protect\subref{fig:transmissionTeleport}~shows quantum teleportation, where entanglement and a classical wireless transmission implements transmission of an arbitrary qubit $\ket{\psi}$. In timeslot $t_2$, Alice and Bob acquire a Bell pair either by generating it directly or requesting it from the 1Q system. Note that Bob first acquires a modified version of the qubit, denoted by $\ket{\tilde\psi}$, which, using the classical information, is finally transformed into the original qubit $\ket{\psi}$.}
 \label{fig:transmission}
\end{figure*}


Classical digital information describes discrete states. Each state can be represented by a digit or a sequence of digits. A bit distinguishes two states. Bits $0/1$ can be universally used to describe digital information. For instance, a data packet is a sequence of bits. In terms of physical information carriers, classical wireless communication typically relies on radio waves, but it can also use Free Space Optical (FSO) transmission, for example employing on/off-keying. 
The general procedure at a classical wireless transmitter is to retrieve the data from some memory or processing unit, map it onto the wireless physical carrier, and send it to the receiver. The role of the designer is to make the modulated waveforms distinguishable by the receiver, despite noise or other channel impairments. \pp{Fig.~\ref{fig:transmission}(a) depicts classical communication, where the sender Alice sends a \emph{copy} of its data to the receiver Bob.} 

\pp{Many things are quite different for quantum information. To start with, the} basic unit is a \emph{qubit} and, for the moment, it is sufficient to say that the state of the qubit is represented by a pair of complex numbers $(\alpha,\beta)$, such that $|\alpha|^2+|\beta|^2=1$. A communication engineer may get an immediate urge to modulate information into $\alpha$ and $\beta$, considering that they are continuous variables and can, theoretically, contain an infinite number of classical bits. Unfortunately, this is not possible, because even if the transmitter modulates information into $(\alpha,\beta)$, the receiver cannot retrieve it. The receiver can retrieve the information stored in a qubit by using a measurement that can have one of the two possible outcomes. We denote those two outcomes by $\ket{0}$ and $\ket{1}$, respectively, just to differentiate them from the classical bits $0$ and $1$; the deeper meaning of those will unfold later. In a compact form, a qubit can be written in the bra-ket notation:
\begin{equation}
    \ket{\psi}=\alpha\ket{0}+\beta\ket{1}
\end{equation}
Upon measurement, the receiver gets $\ket{0}$ with probability $|\alpha|^2$ and $\ket{1}$ with probability $|\beta|^2$. This shows why the urge to modulate information into $(\alpha,\beta)$ is unjustified: in classical settings, this would be equivalent to modulating information into a Bernoulli parameter $p$, where each of the two possible outcomes is sampled from the Bernoulli distribution, but a single observation does not reveal what $p$ is.

There is another subtle difference. Classical information quantifies the uncertainty the receiver has about a state transmitted by a sender. It tacitly assumes that the sender can observe the classical information and does not have the uncertainty that the receiver faces before receiving the data. To elaborate, if Alice prepares the qubit in a state with specific $(\alpha,\beta)$, then she knows the qubit state. However, for arbitrary qubits that are generated by a process or have been left for some time to evolve on their own, \emph{no} observer knows $(\alpha,\beta)$. For instance, in distributed quantum computing, the quantum state obtained at a certain computing step of a distributed quantum algorithm can be unknown. Thus, in quantum communication the sender sends information about the state that is, in general, unknown to the sender. 

The problem with the unknown quantum state is that it cannot be copied, as stated by the no-cloning theorem~\cite{Wootters1982}. \pp{Here \emph{unknown} means that we do not know $\alpha,\beta$ of the qubit and, as explained above, we cannot measure them.} As a consequence, a qubit state cannot be copied from a Quantum Processing Unit (QPU) or a quantum memory to a wireless quantum information carrier, e.g. a photon. The qubit state can exist either in the memory or in the carrier, but not in both. \pp{This is the fundamental difference between classical, Fig.~\ref{fig:transmission}(a), and quantum communication, Fig.~\ref{fig:transmission}(b). In the latter, it is not possible to send a copy of an unknown qubit from Alice to Bob, as only one version of that quantum information can exist at one time. Hence, the transmission of quantum information encoded in a qubit state involves the process of \emph{quantum transduction}, i.e., providing another physical carrier for the same information qubit, see Fig.~\ref{fig:transmission}(b).} \pp{A qubit state realized in a quantum memory is transduced} into a \emph{flying qubit}, to be transmitted. A flying qubit is physically a photon sent over an optical fiber or an FSO link. Quantum transduction~\cite{DavCacCal24,CalDavHan-25} is a critical operation in a distributed/networked quantum system and its impairments corrupting the quantum information should be minimized.

Even after the successful transduction, sending a flying qubit through atmospheric channels exposes the quantum information to irreversible impairments from scattering, absorption, or detector inefficiency. Due to the no-cloning theorem, such loss permanently erases the quantum information. This fragility limits direct qubit transmission to near-perfect or short-range atmospheric conditions, impractical for global-scale quantum networks. 

\pp{The mode of quantum communication considered is a direct transfer of quantum information. A more robust alternative is presented by the use of \emph{quantum teleportation}, but to describe it, we first need to introduce the phenomenon of \emph{quantum entanglement}. A typical example of two entangled qubits is the Bell state $\ket{\Phi^+}=\frac{1}{\sqrt{2}}(\ket{00} + \ket{11})$, also known as an instance of an Einstein-Podolsky-Rosen (EPR) pair of qubits. Entangled qubits exhibit correlations that persist even when the pairs are separated by vast distances. This is a special type of correlation that does not have a classical counterpart.} 

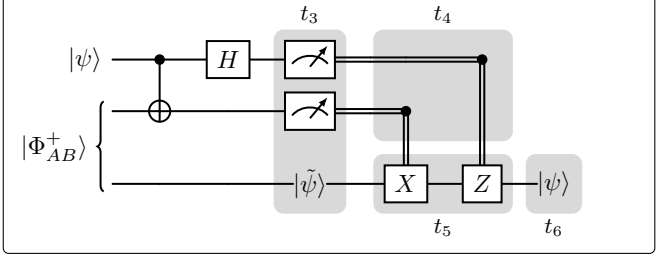
\begin{figure}[bt]
          \centering
          \begin{tikzpicture}
            \draw(0,0) node[anchor=north west](box){
                \begin{minipage}{.95\columnwidth}
                  \color{black}\small
                  Quantum teleportation consists of four steps, as illustrated in Figure~1c and described below.
                  \begin{enumerate}
                    \item Alice sends her part of the entangled EPR pair to Bob at time $t_2$, while retaining the other entangled qubit as well as the information-carrying qubit $\ket{\psi}$.
                    \item At time $t_3$, Alice completes the BSM (Bell-State Measurement) process, by measuring both the qubits in her possession. This results in two measurement outcomes, represented by two classical bits. Note that the effect of the BSM process is that Bob's part of the original entangled pair collapses into a state $\ket{\tilde{\psi}}$, which is an eventually distorted version of the original informational qubit $\ket{\psi}$.
                    \item At time $t_4$, Alice sends these two classical bits to Bob through a classical channel.
                    \item At time $t_5$, Bob uses these two classical bits to determine which of four possible quantum operations should be performed on $\ket{\tilde{\psi}}$. After applying the selected quantum operation, Bob recovers the information qubit $\ket{\psi}$ at time $t_6$.
                  \end{enumerate}
                  Interestingly, the original quantum state is destroyed at Alice's side as a result of the measurement, so no copying occurs. Moreover, the original entangled pair is also destroyed. Thus, any subsequent teleportation requires the generation and distribution of a new EPR pair.
        
                  \medskip
                  \includegraphics[width=.9\columnwidth]{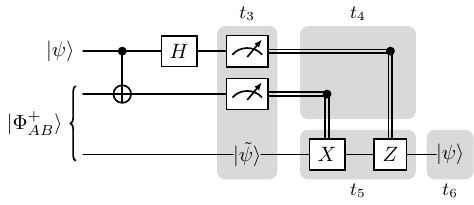}
                  
                \end{minipage}
            };
            \draw[black](box.north west) -- (box.north east) [rounded corners=2pt] -- (box.south east) -- (box.south west) [sharp corners] -- cycle;
            \draw[black, fill=gray!20][rounded corners=2pt](box.north east) -- ++(0,1.5em) -- (0,1.5em) [sharp corners] -- (box.north west);
            \draw(box.north west) node[anchor=south west, font=\bfseries](boxTitle){Quantum teleportation};
          \end{tikzpicture}
          \caption{\pp{Description of quantum teleportation and its associated circuit diagram. Alice holds the two topmost qubits, while Bob holds the bottommost. The labels $t_3$ to $t_6$ correspond to the timeslots given in Figure~1c.}}
                  \label{fig:infoQuantumTeleport}
        \end{figure}

\pp{Assume that Alice generates\footnote{\textcolor{black}{Without loss of generality, we assume an ``at source'' entanglement generation scheme \cite{cacciapuoti2020entanglement}. The same assumption is also used in Fig.~\ref{fig:infoQuantumTeleport}}.} an EPR pair $\ket{\Phi^+}$ and sends one of the qubits to Bob. This transmission does not carry any information, since Alice does not map any information onto the qubit sent to Bob. Now, if Alice measures her own qubit from $\ket{\Phi^+}$, the resultant outcome is random, either $\ket{0}$ or $\ket{1}$, each with probability $\frac{1}{2}$. Due to the entanglement, if Bob measures his qubit of the EPR pair, he will get exactly the same outcome as Alice; this is the specific correlation mentioned above. However, no information is sent through this process from Alice to Bob, since Alice has no control over the outcome in her measurement. Furthermore, sending one of the entangled qubits from a pair in a known state, such as $\ket{\Phi^+}$, is different from the transmission of a qubit $\ket{\psi}$ in an unknown state. While the latter cannot be copied, Alice can, in principle, create an arbitrary number of copies of the known Bell pair $\ket{\Phi^+}$. Say that Alice sends an entangled qubit from $\ket{\Phi^+}$ to Bob and, during its transmission, it gets lost or loses entanglement with the qubit left at Alice, owing to the environmental contamination. Then this qubit is perfectly replaceable by another qubit from another Bell pair $\ket{\Phi^+}$. Hence, unlike the unknown qubit state $\ket{\psi}$, an entangled qubit is \emph{expendable} and can be perfectly replaced with an entangled qubit from another pair; just like in retransmission of classical information in ARQ protocols. Although transduction is also required when different physical systems are involved, entanglement distribution is generally more reliable than direct transmission of an information-carrying flying qubit.}

\pp{At this point, we have covered the concepts necessary to introduce quantum teleportation, a quantum communication protocol that allows the sharing of an informational qubit without the physical transfer of the particle encoding the qubit state. We give the details in Fig.~\ref{fig:infoQuantumTeleport}.}
\pp{Quantum teleportation demonstrates that quantum information can be transferred by leveraging entanglement together with classical communication, thereby overcoming the fundamental limitations that quantum mechanics imposes on the direct transmission of informational qubits. Consequently, the design of a quantum wireless system, such as 1Q, relies on both classical bits and entanglement distribution. The novel functionality of the wireless link, compared to classical wireless mobile networks, is the FSO-based entanglement distribution, pivotal for long-range quantum networks. Future research has to explore whether THz RF links could be realistically harnessed for entanglement distribution, despite their less pronounced particle-like quantum-behavior. }

\begin{figure*}[t!]
  \centering
  \includegraphics[page=4]{qbs.pdf}\\[4mm]
  \subfloat[]{
      \includegraphics[page=1]{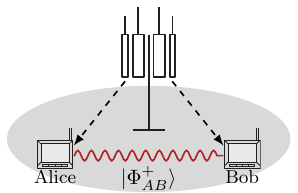}
  }
  \hspace{2mm}
  \subfloat[]{
      \includegraphics[page=2]{qbs.pdf}
  }
  \hspace{2mm}
  \subfloat[]{
      \includegraphics[page=3]{qbs.pdf}
  }
 \caption{Basic functionalities of a Quantum Base Station (QBS). (a) \emph{Quantum wireless access:} 
 Creation of entangled pairs and distribution to devices within its coverage area. (b) \emph{Quantum analogy of classical broadcast:} Creation and distribution of multi-partite entanglements within its coverage area. (c) \emph{Quantum forwarding:} Entanglement swapping between a device within the coverage area and a quantum node connected to the QI.}
 \label{fig:QBSfunctionality}
\end{figure*}

\section{1Q System Model}
\label{sec:1Qsystemmodel}

\pp{This section describes the key architectural elements of the 1Q system: base stations and cells. Specifically, we denote the traditional base station for classical information as \gls{cbs} and introduce \gls{qbs} that brings in the quantum internet functionality.}

\subsection{Functionality of an 1Q Base Station}

The basic functionality of a \emph{\gls{cbs}} is summarized as: 
\begin{enumerate}
    \item \emph{Wireless downlink/uplink access.} A \gls{cbs} offers wireless network access to devices within its coverage area, that is, downlink and uplink transmission of data and metadata or  control information.
    \item \emph{Broadcast.} A \gls{cbs} can exploit the broadcast nature of the classical wireless links and reach multiple receivers in the downlink with a single transmission. This is of fundamental importance for managing the devices within the  coverage region of a \gls{cbs}, such as advertising the presence of \gls{cbs} for initial access or support of the mobility management.
    \item \emph{Forwarding.} A \gls{cbs} that provides network access is almost never the ultimate communication peer of the devices within its coverage. Explicitly, the \gls{cbs} forwards information towards the core network and the global Internet. 
    \item \emph{Mobility support.} A \gls{cbs} takes part in handover procedures, through coordination with the core network and other \glspl{cbs}, in order to offer seamless connectivity for users whose wireless coverage changes over time.
    \item \emph{Computing, sensing, positioning.} In 5G the \glspl{cbs} also offers computing resources, as in edge computing that supports devices locally, within the \gls{cbs} coverage. The \gls{cbs} functionality is further broadened towards 6G by including sensing and positioning. 
\end{enumerate}
These functions must be accomplished while maintaining both timing and protocol coordination, and operating within the available spectrum, time, and power constraints of the system.

Similarly, we can define the functionality of a \emph{\gls{qbs}}. A QBS can offer wireless access for quantum information exchange for the devices within its quantum wireless coverage, that is, FSO coverage. Instead of direct qubit transmission, QBSs should provide wireless access through quantum teleportation. This implies that QBSs should be able to distribute entanglement within their coverage and provide classical wireless access.  

Regarding the second CBS functionality, due to the no-cloning theorem, QBSs cannot broadcast arbitrary quantum information to multiple receivers. As an analogy, instead of the classical broadcast property, we will assume that QBSs can distribute \emph{any} entanglement to the devices within their coverage region. We have previously introduced the pairwise entanglement for a Bell pair $\ket{\Phi^+_{AB}}$. Here the subscript $AB$ means that one of the entangled qubits is at $A$ and the other at $B$. The QBS can create the entangled pair $\ket{\Phi^+_{AB}}$ and then send one entangled qubit to each of the two devices within its quantum coverage,
as shown in Fig.~\ref{fig:QBSfunctionality}(a). More generally, QBSs can create and distribute entanglement across more than two parties, known as \emph{multipartite entanglement}. For instance, QBSs can create a 3-qubit Greenberger-Horne-Zeilinger~(GHZ) state~\cite{Greenberger1989}:
\begin{equation}
    \ket{\mathrm{GHZ_3}}=\frac{1}{\sqrt{2}}\left(\ket{0_Q0_A0_B}+\ket{1_Q1_A1_B}\right).
\end{equation}
The QBS keeps the first entangled qubit, while sending the second to Alice and the third to Bob, as in Fig.~\ref{fig:QBSfunctionality}(b). This state 
enables the QBS to realize a certain distributed quantum computing task in concert with two devices within its coverage. More generally, QBSs can create and distribute arbitrary multipartite entanglements involving $N$ qubits within their quantum coverage region. 

Regarding forwarding, a QBS should act as a \emph{quantum repeater} for long-distance quantum communication between the devices within its coverage and quantum nodes connected to the global QI. Long-distance entanglement, needed in teleportation, is provided by using \emph{entanglement swapping}, a quantum protocol in which two particles that have never interacted (each initially entangled with separate particles) become entangled by performing a joint measurement at an intermediate node. QBSs should be capable of performing entanglement swapping  and, in this way, ensure long-distance entanglements for the devices within their coverage. This is shown on Fig.~\ref{fig:QBSfunctionality}(c). Clearly, QBSs can also perform entanglement swapping for two devices within their coverage.   

The distinctive feature of the mobility support in a quantum context is related to entanglement distribution. Considering the short coherence times, entanglement distribution should not only be timed based on the spatial location of the user, but also the timing of the required service. \pp{This can be different, depending on whether the \gls{qbs} is fixed, terrestrial, or a mobile \gls{qbs} on a satellite.}
When the entanglement is distributed by a QBS placed in a Low Earth Orbit (LEO) satellite constellation, it can then take advantage of the predictable handover time in addition to the service request from the users.

Considering the fifth functionality, a QBS can facilitate a local distributed quantum computation within its coverage area. In other words, a QBS can offer the wireless variant of a \gls{qlan}~\cite{alshowkan2021reconfigurable}, a network that interconnects  quantum computing devices, quantum memories, and quantum sensors within a localized physical area. 

\subsection{Classical and Quantum Cells}
\label{sec:class-quant-cells}

Having contrasted the functionality of \gls{qbs} to \gls{cbs}, we define two different cell types, classical and quantum, respectively. 

A \emph{classical cell}, or shortly \emph{cell}, is the area covered by a \gls{cbs}, so that any device within that area can transmit or receive classical information to or from that \gls{cbs}. The physical carrier of information in the classical cell is assumed to be a radio wave, although classical information can also be transmitted using FSO links. By contrast, a \emph{quantum cell} is covered by a \gls{qbs}, so that: (1) The \gls{qbs} can distribute entanglement among two or more nodes within this area. (2) Any device within the area can exchange quantum information with the \gls{qbs}. Photons are assumed to be the physical carriers of qubits, so that wireless quantum communication within the quantum cell is based on FSO links.

The \gls{cbs} and \gls{qbs} can be collocated within the same infrastructure node; we refer to it as \gls{bs}, with both classical and quantum coverage. A mobile device that has both classical and quantum capability is denoted as \gls{que}, while a device without quantum capability as \gls{cue}. 

The coverage of the classical and quantum cell is, in general, different due to the propagation difference between radio and light waves. The classical and quantum information also different integrity, where the latter suffers from decoherence. In this regard, experiments show that terrestrial FSO links in urbanized environments can distribute entanglement over ranges of $1.5$~km~\cite{Nafria:23}, which is comparable to the coverage ranges in classical urban cells. Satellite FSO links can distribute qubits over distances of hundreds of kilometers~\cite{yin2017satellite}. Their range can also be further enhanced by optical reconfigurable intelligent surfaces (ORISs) in space-air-ground integrated networks (SAGIN)~\cite{trinh2025ORIS}. 

From a system-oriented perspective, each device that has a quantum coverage must also have a classical coverage, but not vice versa. This is because classical communication is necessary to send metadata about the quantum states and measurements. In principle, the FSO used for quantum information can also transmit photons conveying classical information. Transmission of both bits and qubits over the same physical carrier in optical fiber has given rise to problems of coexistence. While coexisting quantum and classical communication can be achieved at distances of up to $100$~km, the typical classical communication rate are extremely low, in the order of $2.3$~kbps in quantum secure direct communication (QSDC)~\cite{Pan25}. In the C band, however, high QKD rates of up to $1.3$~kbps have been achieved while transmitting classical data at up to $1.6$~Tbps~\cite{Schreier23}. In this initial discussion we assume separated physical carriers for classical and quantum information, respectively. 

A \gls{bs}, as an infrastructure node, can be static or dynamic. For instance, a terrestrial \gls{cbs} or \gls{qbs} is static and its coverage, generally, does not change over time. Nevertheless, there can be unplanned fluctuations in the coverage due to movements in the environment and, e.g., intermittent occurrence of obstacles. A \gls{qbs} mounted on a satellite is a dynamic infrastructure node, as its coverage changes over time in a planned and, often, predictable way. 

\glspl{cue} do not participate in exchanges of qubits, but they may benefit from, e.g., quantum-based cloud processors. Yet, for the purpose of 1Q, we will view them as an additional load on the classical time-frequency-antenna resource grid. 
The performance of \glspl{cue} is typically quantified using link- and system-level metrics such as data rate, reliability, or latency-related measures. Instead, the performance of \glspl{que} is characterized by quantum-specific measures of the quantity and quality of quantum resources available for advanced applications. Quantity metrics include the entanglement generation rate and the secret-key rate for \gls{qkd} scenarios, while quality metrics encompass the fidelity or quantum bit-error rate (QBER). 

\begin{figure}[t!]
 \centering
 \includegraphics[width=0.9\columnwidth]{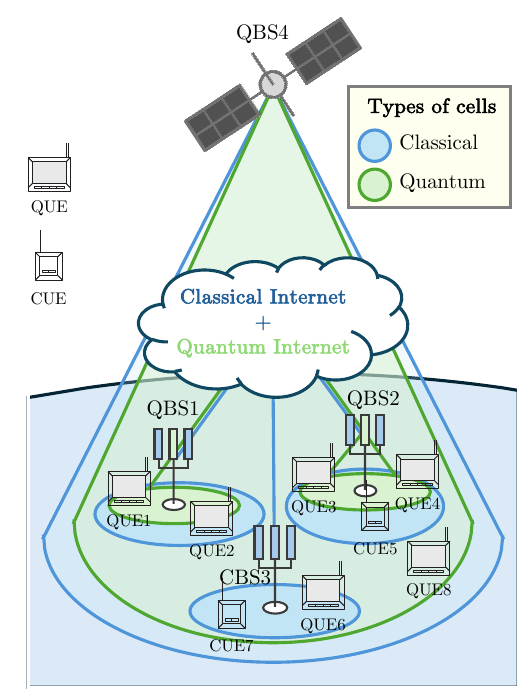}
 \caption{A primer of a 1Q system with classical/quantum base stations, cells, and \glspl{ue}, including both terrestrial and non-terrestrial coverage.} 
 \label{fig:1Qsystem}
\end{figure}

Fig.~\ref{fig:1Qsystem} exemplifies an infrastructure having quantum and classical cells. The classical/quantum cells defined by the satellite \gls{qbs} have a large coverage, but offer limited  service rates, namely classical data rate or number of entanglements per second. The terrestrial base stations have lower coverages, but potentially higher service rates.
Terrestrial and satellite \gls{bs}s must operate in unison to provide truly ubiquitous coverage seamlessly spanning indoor and outdoor, rural and urban areas. 

We illustrate the operation using examples from Fig.~\ref{fig:1Qsystem}:
\begin{itemize}
    \item QUE$_3$ and QUE$_4$ are in the classical and quantum coverage of QBS$_2$. This means that QBS$_2$ can \emph{reactively} distribute entanglement to QUE$_3$ and QUE$_4$ upon receiving requests from these devices that relate to a certain quantum task, such as computing or sensing. 
    Furthermore, QUE$_3$ and QUE$_4$ compete with CUE$_5$ to use the classical radio resources.
    \item QUE$_1$ is in the quantum/classical coverage of QBS$_1$ and can receive both bits and qubits, while QUE$_2$ can only communicate classically with QBS$_1$. However, QBS$_1$ may have \emph{proactively} distributed entanglement to QUE$_2$, while it was located within the quantum coverage. Under sufficiently long coherence-time, QUE$_2$ can still send/receive quantum information through teleportation. 
    \item QUE$_3$ and QUE$_4$ can also use the entangled qubits for secure communication. However, it is not likely that the communicating device will be in the same quantum cell. Here QBS$_2$ should enable entanglement swapping and let QUE$_3$ acquire entangled qubits with devices that are outside the coverage of QBS$_2$. This is exactly the same entanglement distribution mechanism across QLANs, as described in~\cite{Mazza:25}. It can establish entanglement between, say, QUE$_1$ and QUE$_3$ without using the satellite QBS$_4$. 
    \item Let the satellite QBS$_4$ distributes multipartite entanglement to QUE$_1$, QUE$_6$, and QUE$_8$. Following the QI approach based on Local Operations and Classical Communication (LOCC), the multipartite entanglement can be used as a basis to create pairwise entanglement between, say, QUE$_6$, and QUE$_8$, using only the classical terrestrial coverage of QUE$_6$.  
\end{itemize}

\subsection{Physical Constraints and Spectrum}

Designing a networking system can, to a large extent, proceed by treating information, classical or quantum, as an abstract entity. Nevertheless, all information is eventually physical~\cite{landauer1991information} and here we discuss two physical aspects that are relevant for the system design: (1) wireless spectrum and (2) decoherence and volatility of the quantum resources. 

\subsubsection{Wireless Spectrum}
Wireless classical communication relies upon the radio spectrum and, with the emergence of 5G/6G systems, there has been a tendency towards higher radio frequencies and greater bandwidths. Classical cellular networks have defined three frequency bands, known as Frequency Ranges (FR) FR1 (< $6$ GHz), FR2 (30-100 GHz), and FR3 (located between FR1 and FR2). As indicated previously, wireless quantum information is sent through flying qubits and entanglement distribution that rely upon \gls{fso}. Thus, besides the qualitative change, quantum communication quantitatively expands the use of the wireless spectrum towards higher frequencies, suitable for transmission of qubits. This suggests the addition of a new band, FR4, in the peta-Hertz (PHz) range, for the transmission of data via laser beams in \gls{los} connections. The need for \gls{los} highlights the strategic advantage of satellite \glspl{qbs}, which could be complemented by ORISs  deployed on building rooftops~\cite{trinh2025ORIS}.

\subsubsection{Decoherence, Fidelity, and Timing Constraints}

Quantum algorithms and applications rely upon the ``quantum behavior'' of the physical resources, including superposition and entanglement. Decoherence is the process whereby the quantum behavior erodes owing to interactions with its environment. 
Closely related to this, the fidelity $F$ of a qubit quantifies the similarity between the quantum state $\rho$ of the qubit in question and the state of an entangled qubit $\varphi_{ideal}$. Here $F = \bra{\varphi_{ideal}}\rho\ket{\varphi_{ideal}}, 0 \leq F\leq 1$.

For system designers, decoherence translates into volatility and unreliability of quantum resources, affecting the performance and dependability of distributed quantum systems in computing, sensing, and security. Specifically, a certain quantum task must be finalized within the coherence-time, before the quantum resources decohere. This imposes \emph{quantum timing constraints} on the system design. On the other hand, the quantum task, be it computing or sensing, is a component of another digital computing application; for instance, quantum search done as a part of a decision process. The overall digital application may have another real-time constraint, termed \emph{digital timing constraint}, requiring finalization of the decision process within a certain deadline, say imposed by a robotic system. 

\begin{figure}[t!]
 \centering
 \includegraphics[width=8.3cm]{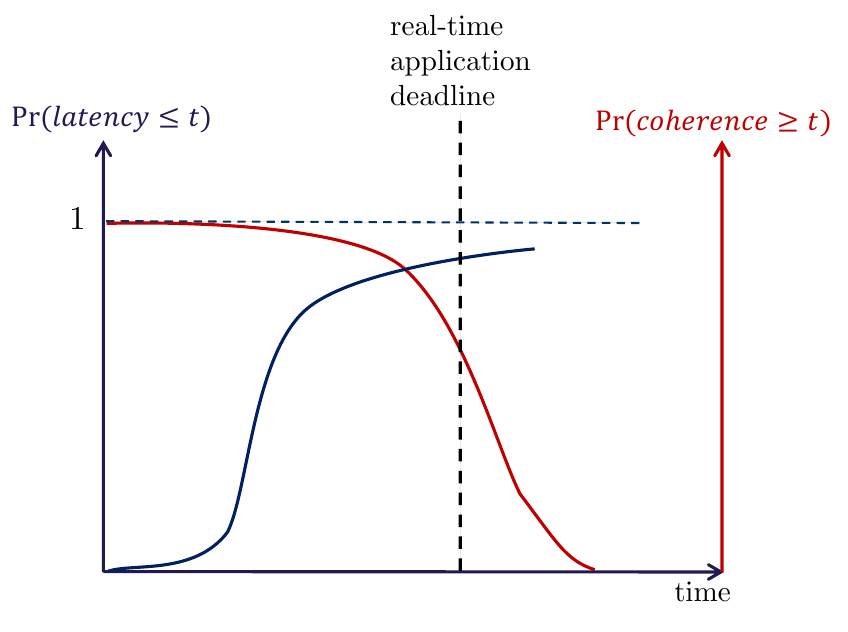}
 \caption{Conceptual depiction of timing requirements, digital due to application and quantum due to decoherence.} 
 \label{fig:DigitalQuantumTiming}
\end{figure}

This is illustrated on Fig.~\ref{fig:DigitalQuantumTiming}. Digital timing has to meet the requirement by which a task should be finalized within a certain real-time deadline with reliability at least $1-P_e$. In the simplest case, this task can be successful transmission of a classical data-packet. If the computing resources are not volatile, as in classical systems, the probability that the task will be successfully fulfilled increases over time. More formally, the probability that the latency of the task execution is lower than some time $t$ increases as $t$ grows. This is because, as time passes, the system can do more to increase the reliability: e.g., the data-packet can be retransmitted if there is an error, so that the overall probability of success will increase at the expense of higher latency. This has exploited by the Ultra-Reliable Low Latency Communication (URLLC) service in 5G~\cite{popovski2019wireless}.

In parallel, Fig.~\ref{fig:DigitalQuantumTiming} shows the impact of the quantum timing constraints. The probability that the quantum resources will stay coherent beyond some time $t$ decreases as $t$ increases. This tendency challenges the logic of the other curve, where accuracy/reliability improves over time. The consequence for system design is that the quantum resources have to be created and used in concert with the dynamics of the overall digital application. For instance, entanglement may be distributed just-in-time, before it is exploited. Moreover, the procedures of improving entanglement distribution, such as entanglement distillation~\cite{rozpkedek2018optimizing}, must be factored in through their timing-reliability characteristics.

\section{1Q: Application Scenarios}
\label{sec:ApplicationScenarios}

\gls{qkd} is currently the most mature quantum application. Another application, \gls{qsdc}, provides security over entanglement-based networks without setting up a private key session. 1Q could support QSDC protocols, leveraging hundreds of kilometers of practical optical fiber while allowing its integration with satellites using FSO and quasi-optical/THz radio-frequency~\cite{li2023single}.

In this section we turn attention to two less established scenarios relevant for 1Q, which are quantum computation and quantum sensing. 
\subsection{Quantum Computing}
At present, the available \glspl{que} are Noisy Intermediate-Scale Quantum (NISQ) devices, having memories constrained to the order of $100$ physical qubits. On the other hand, producing a fault-tolerant logical qubit may require $10^3$--$10^4$ physical qubits. The communication procedure in the distribution of qubits from the \gls{qbs} to the \glspl{que} for distributed quantum computing is  limited by the available NISQ processors. Hence, 1Q will likely rely upon a hybrid quantum-classical computing framework. Nevertheless, the use of FSO channels in 1Q networks is challenged by quantum decoherence, pointing errors, and propagation impairments over long distances, as well as synchronization impairments~\cite{trinh2024quantum}. 

We discuss two quantum computing instances. 

\begin{figure}[t!]
 \centering
 \includegraphics[width=7cm]{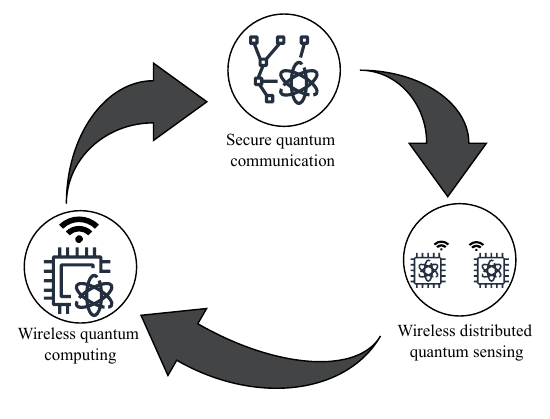}
 \caption{Three generic application scenarios in 1Q: (1) Secure Quantum Communication, (2) Wireless Quantum Computing, and (3) Wireless Distributed Quantum Sensing.}
 \label{fig:usecase}
\end{figure}

\subsubsection{Distributed Quantum Computing}  1Q can interconnect \glspl{qbs} and \glspl{que} and offer the functionality of a \gls{qlan}~\cite{alshowkan2021reconfigurable,CacCalIll-25}, a concept motivated by the limitations in the number of qubits in NISQ processors. In a small-scale \gls{qlan}, the interconnection of multiple Quantum Processing Units (QPUs) would support a distributed quantum computation architecture. However, 1Q has to deal with the heterogeneity of interconnections and hardware; see \cite{CacCalIll-25} for distributed quantum computing types: multicore, multicomputer, and multifarm. 
In this context, communication protocols should support: \emph{(i)} interconnection of quantum computing modules that are geographically separated, and \emph{(ii)} inter-processor operations, e.g. when a quantum gate must be applied to a remote qubit in a multi-core distributed archtecture. This is where classical networks can augment their quantum counterparts: leveraging classical bits to supplement the exchange of quantum information, such as classical bits encoding measurement outcomes of the qubits during the quantum teleportation operation. 
In a 1Q network, \gls{qbs} would serve as an anchor to support on-demand Entanglement-as-a-Service (EaaS) for classical-quantum information processing in quantum applications and services. For example, classical data can be encoded into specialized quantum states in a quantum cell to perform communication-efficient distributed computations, allowing quantum speedups for complex problems in classical networks \cite{gilboa2024exponential}. However, there is a non-negligible cost of embedding classical data into quantum states for quantum processing.

\subsubsection{Quantum Machine Learning (ML) and Optimization} 

The 1Q network architecture enables \glspl{que} to have wireless access to quantum ML resources for gleaning quantum advantages through efficient training procedures and accelerated inference. First, the classical data at \glspl{que} should be turned into quantum data, and this conversion requires computational resources. Therein, the \glspl{que} can exchange classical data (local quantum models) to train high-quality quantum models such as variational quantum classifiers (VQCs). Once trained, quantum models provide prompt inference by harnessing fewer operations required by VQCs than their classical counterparts. Since \glspl{que} are NISQ devices having hardware limitations, the \glspl{que} can, in principle, access quantum computing resources available at \gls{qbs} for directly processing the local data, sent via classical channels, for quantum model training at the edge. Specifically, the compute-intensive quantum kernel estimation procedure required to map data into high-dimensional Hilbert spaces and to model training thereafter are offloaded to the quantum server residing at the \gls{qbs}. The same operations can be executed securely with protocols such as Blind Quantum Computing (BQC)~\cite{UniversalBQC}, where the remote quantum server remains oblivious to the content of the delegated  computation.

There is a potential for an exponential quantum advantage in communication using distributed \glspl{que} across multiple quantum cells, as in 1Q networks, for distributed learning; however, it is restricted by hardware constraints of simple circuits~\cite{gilboa2024exponential}. 
The speed-up attained by the quantum algorithms over classical algorithms, such as search and optimization, simulation of quantum systems and learning, should dominate the time required for encoding data into quantum states through efficient design of quantum circuits~\cite{caleffi2024distributed}.

\subsection{Wireless Distributed Quantum Sensing} 
Quantum sensing enables high-precision measurements using entangled systems. The 1Q networks establish an architectural premise with entangled systems for wireless distributed sensing. QUEs, equipped with quantum sensors~\cite{degen2017quantum}, participate in networked quantum sensing and measure the properties of a physical quantity~\cite{degen2017quantum}. Therein, a QBS acts as a source to create entangled probe states shared by the QUEs for jointly interrogating the quantity.
In principle, the \gls{qbs} can generate a multipartite entangled probe state, which can then be shared among $N$ \glspl{que} through FSO transmissions to scan an object of interest.
By contrast, distributed classical sensing can be interpreted as using a separable probe state.
In either case, the measurements can be post-processed \emph{classically}: once the measurements are done, classical communication channels are used to transmit the measurements to a central node (QBS) for post-processing. Classical communication is also necessary for coordinating and consolidating the final measurement output of the sensed object. Furthermore, 1Q will use classical signals for time synchronization among \glspl{que} and control feedback loops for error correction; thereby, playing an important role to support the distributed quantum sensing process. Repeated measurements, i.e., taking $N$ measurements from independent \glspl{que} and averaging the individual results provides sensitivity improvements proportional to $1/\sqrt{N}$, commonly referred to as the standard quantum limit (SQL).

Noise poses a fundamental challenge in such operations: there exists a nonzero error in experimental detection of measurable transitional probability that indicates the change in the state of qubits at each \glspl{que}. In 1Q networks, classical readout noise would exacerbate the other existing sources of uncertainties and errors, such as quantum projection noise, decoherence and relaxation in sensing time, plus imperfect initialization of the quantum sensor. Therefore, the innate understanding of the noise spectrum and design of suitable sensing protocols are crucial for defining performance limits and sensitivity (sensing precision) in distributed quantum sensing and its mapping to the 1Q protocols.

\section{Illustrative Examples of 1Q Use Cases}
\label{sec:QKD_examples}
In general, the timeline of quantum applications, including those in 1Q, consists of four phases. (1) \emph{Setup,}  typically initiated as a service request from a \gls{que} to a \gls{qbs}. (2) \emph{Entanglement distribution,} in which the \gls{qbs} generates entanglement and distributes it to the relevant \gls{que}s. (3) \emph{Local quantum operations,}  e.g., qubit measurements or gates. (4) \emph{Reconciliation,} using classical operations, particularly communication of measurement outcomes between Alice and Bob, but also post-processing. Phases (1) and (2) correspond to resource management, while the combination of (3) and (4) is the execution phase. The execution phase is analogous to the operation of distributed computing architectures, where any computing task can be decomposed into blocks of computation followed by communication of the task data and synchronization. Some applications, though, may require several rounds of the execution phase.

First, assume that a given quantum application is requested at time $t_0$ with \emph{digital time} budget $T_d$, and coherence or \emph{quantum time} budget $T_q$. It is tacitly assumed that these times contain a statistical quantifier: for instance, $T_q$ is the time for which the fidelity is above a critical threshold. Hence with a certain predefined probability, we regard the application as successfully concluded, provided that the lifetime (time until decoherence) of every qubit is within $T_q$, and that the end-to-end execution, from service request through processing, is completed within $T_d$. Otherwise, the application is deemed unsuccessful if the lifetime of any qubit is below $T_q$, or the time budget $T_d$ is exceeded. 

All aforementioned applications require reliable transfer of quantum and classical information, regardless whether across nodes in a network (communication), between memory units (computing), or from a physical system to a readout (sensing). Thus,  a generalized expression for the performance of any of these applications in terms of the probability of successful execution integrates both the effective error probability associated with: 1) the quantum layer $P_{\text{err}}^{(q)}$  (e.g., loss, decoherence, low gate fidelity, and state preparation error), and  2) the classical imperfections $P_{\text{err}}^{(c)}$ (bit flips in control electronics, timing jitter, and synchronization errors). Note that $P_{\text{err}}^{(q)}$ and $P_{\text{err}}^{(c)}$ are functions whose parameters depend on the specific application. For instance, when running a quantum communication protocol, $P_{\text{err}}^{(q)}$ might capture the effects of photon loss and channel noise, while $P_{\text{err}}^{(c)}$ captures synchronization errors and imperfect reconciliation algorithms. On the other hand, during an instance of quantum computation, $P_{\text{err}}^{(q)}$ is determined by quantum gate and memory errors, whereas $P_{\text{err}}^{(c)}$ includes faulty syndrome decoding, control electronics noise, or misclassification of measurement outcomes. Despite these differences, this formulation highlights a structural symmetry: regardless of whether the protocol is used to distribute entanglement, execute an algorithm, or sense a weak field, the overall success probability hinges on the joint reliability of the quantum and classical channels. By treating these as layered but interdependent processes, we can directly compare different quantum technologies within a unified performance framework.

Suppose a given protocol consists of $K$ blocks of operations, both quantum and classical, as indicated for QKD in Fig.~\ref{fig:qkdTiming}, and let $P_{\text{succ}}(i)$ denote the success probability of the $i$'th block in the protocol. Then, the global success probability is
\begin{align*}
    P_{\text{succ}} 
    &= \sum_{i=1}^K \left(P_{\text{succ}}(i)\prod_{\ell=1}^{i-1}\Big(1-P_{\text{succ}}(\ell)\Big)\right).
\end{align*}

Consider the simple case where 
\begin{equation*}
    P_{\text{succ}}(i)=p=\left(1-P_\text{err}^{(q)}\right)\left(1-P_\text{err}^{(c)}\right)
\end{equation*} for all $i=1,\ldots,K$. Then the global probability of success is
\begin{align*}
    P_{\text{succ}} 
    &= 1 - (1-p)^K \\
    &= 1 - \left((1-P_{\text{err}}^{(c)})P_{\text{err}}^{(q)}+P_{\text{err}}^{(c)}\right)^K.
\end{align*}

Successful completion of each of the quantum and classical operations within the blocks might require multiple attempts due to errors. Furthermore, redundancy might be added to the protocol to combat potential errors in specific operations, e.g., by distributing additional qubits or classical bits, or simply by repeating the activation of certain blocks that presented unrecoverable errors. Consequently, the time to complete each block is a random variable and, thus, the execution time of the application is also a random variable, which might eventually exceed $T_d$. Therefore, as in classical communication applications, reliably achieving a total execution time below $T_d$ requires a delicate balance between the timing and reliability of each of the quantum and classical operations in the protocol for the specific application. To illustrate the operations of the 1Q system within this framework, we present an application for each of the three generic use cases from Fig.~\ref{fig:usecase}.

\subsection{Quantum Key Distribution (QKD)}
The aim of QKD is to establish a shared, common secret between two devices. To illustrate the operation of 1Q, we use the protocol BBM92~\cite{bbm92} which relies upon entanglement. In BBM92, two devices request entangled qubit pairs in a specific state from a given source, which can be trusted or untrusted, and one photon is sent to each of the two devices.  Explicitly, assume that the state is $\ket{\Phi^+}=(\ket{00}+\ket{11})/\sqrt{2}$, which may also be written as $\ket{\Phi^+}=(\ket{{+}{+}}+\ket{{-}{-}})/\sqrt{2}$ with $\ket{{\pm}}=(\ket{0}\pm\ket{1})/\sqrt{2}$.
For each of these pairs, the two devices individually measure their qubit, with each measurement performed in a basis chosen uniformly between $\{\ket{0},\ket{1}\}$ and $\{\ket{{+}},\ket{{-}}\}$.
Note that once a photon is shared, it becomes subject to decoherence. In this specific case, each qubit can be measured immediately after reception, and the resultant qubit can be discarded since only the classical measurement result must be stored.
Therefore, the BBM92 protocol is not as susceptible to issues caused by decoherence as other protocols requiring extended storage time in quantum memories.

By construction, those pairs where the devices choose the same basis will always yield identical measurement results at both devices.
Thus, after a pre-defined number of qubits are distributed and measured, the devices simply have to reconcile/discuss their choices of bases, meaning that they will openly share these choices.
Bits obtained from disagreeing bases are discarded, and the devices are left with a sequence of bits inferred from the identical bases. Note that, since only two measurement bases are used in the BBM92 protocol, the probability of discarding a qubit, representing the QBER, at this stage is exactly $P_\text{err}^{(q)}=0.5$. The number of qubits measured with the same basis is a binomial random variable 
with $N$ being the number of entangled qubit pairs that are distributed to the QUEs.
Thus, the success probability of QKD for the BBM92 protocol can be calculated as the probability of receiving and measuring $K$-out-of-$N$ entangled qubit pairs with the same basis in both QUEs.

The resultant bit-sequence relying on identical bases represents the {\em sifted key}. Based on the above steps Alice and Bob acquire a highly correlated key-pair, but they might still differ slightly due to either channel-effects or perturbations by the eavesdropper (Eve), who can only use another random basis for her measurements. To estimate the resulting error rate, a subset of the remaining key-bits are sacrificed by publicly announcing their measurement results and checking for disagreement. While a low error rate may be corrected by the classical-domain reconciliation process, an error rate above a certain threshold indicates that the link is compromised, and the QKD protocol is aborted or restarted. As a final step of {\em classical post-processing privacy amplification} is used, which discards some of the key-bits to further confuse Eve.

\begin{figure}
  \centering
  \includegraphics[width=\linewidth]{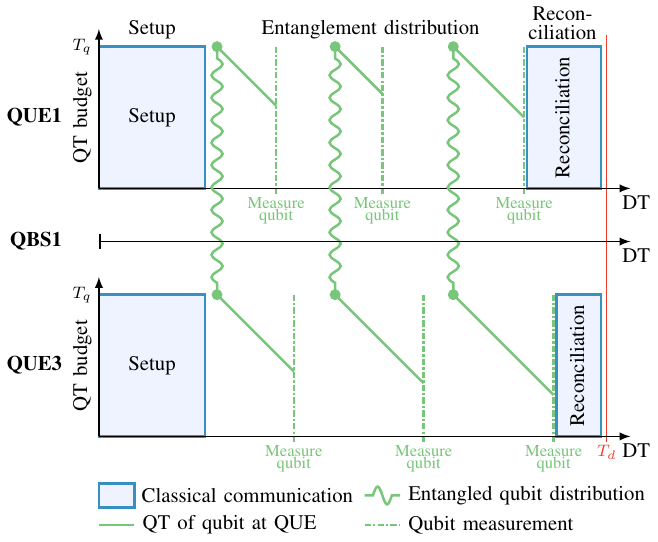}
  \caption{Illustration of the digital and quantum timing budgets (DT and QT) in the case of QKD where QBS1 distributes 3 entangled qubit pairs to QUE1 and QUE3 from Fig.~\ref{fig:1Qsystem}.The horizontal axis indicates digital time and the vertical axis indicates quantum time in the form of remaining time before decoherence.}
  \label{fig:qkdTiming}
\end{figure}

In settings where failed QKD attempts cause devices to retry, the number of qubits exchanged in each round directly affects the end-to-end delay. Namely, longer block lengths, determined by both $K$ and $N$, cause longer delays and more waste of power in case the exchanged bits must be discarded. On the other hand, very short block lengths result in a larger variance in the number qubits measured in matching bases, hence the specific selection of $N$ becomes more critical. This suggests the existence of a 'sweet-spot' for block length when delay is included in the analysis of QKD protocols.

We now concretize these considerations in the scenarios outlined at the end of Section~\ref{sec:class-quant-cells}. QKD protocols can operate in most use-cases illustrated in Fig.~\ref{fig:1Qsystem}. Naturally, the BBM92 protocol can be executed fully with QUEs within coverage of either a single QBS on ground and in space, such as with QUE3 and QUE4 using QBS2 or with QUE3 and any other QUE using QBS4. Here, the common QBS can directly supply the entangled pairs, meaning that quantum timing is only affected by the generation, wireless transmission, and measurement latencies. If at least one of the QUEs is outside the coverage of the QBS generating entangled pairs, these must be transmitted via quantum repeaters over the 1Q network to the QUEs, which increases the risk of decoherence. This process is illustrated in Fig.~\ref{fig:qkdTiming} for the case where QBS1 distributes 3 entangled qubit pairs to QUE1 and QUE3 from Fig.~\ref{fig:1Qsystem}. As QUE3 is outside the quantum coverage of QBS1, the risk of decoherence for QUE3 is higher than that for QUE1. Finally, QKD can be partially executed by distributing and measuring the entangled qubit pairs when the QUEs are within coverage, but the announcement of the measurements can be done afterwards, as long as the QUEs are within the classical coverage of either a terrestrial BS or the satellite QBS4. 

\subsection{Blind Quantum Computing (BQC)}

One of the relevant applications of wireless quantum computing in 1Q is BQC, which enables a quantum device having limited quantum capabilities to delegate a quantum computation to a powerful server, while keeping the computation blind, i.e., private concerning the server. In 1Q, this allows any \gls{que} to access a universal computation run by a server, e.g. at the \gls{qbs}, while maintaining privacy and confidentiality. Several BQC schemes have been proposed harnessing diverse assumptions regarding the capabilities of the delegating device. Here, we focus on the entanglement-based protocol of Universal BQC~\cite{UniversalBQC} that relies on measurement-based quantum computing.

The protocol is initialized by a \gls{que} requesting a \gls{qbs} to generate $K$ EPR-pairs. Then half of each EPR-pair is sent back to the QUE and the other half to the server. For each of these EPR-pairs, the \gls{que} picks an angle $\theta_n\in\{0,\pi/4,\ldots,7\pi/4\}$ and measures its qubit in the corresponding equatorial basis $\{\ket{+_{\theta_n}},\ket{-_{\theta_n}}\}$ with $\ket{\pm_{\theta_n}}=(\ket{0}\pm e^{i\theta_n}\ket{1})/\sqrt{2}$. As a result, the corresponding qubit at the server is remotely prepared in the state $\ket{\pm_{-\theta_n}}$, where the sign depends on the measurement outcome. After all of its qubits have been remotely prepared, the server entangles them into a fixed graph state, typically referred to as a brickwork state due its structure resembling a wall made of bricks. The actual computation then proceeds with sequential single-qubit measurements of the entangled qubits. For each qubit, the \gls{que} transmit the physical equatorial measurement angle $\delta_n=\phi_n-\theta_n+\pi r_n$, where $\phi_n$ is the logical angle and $r_n\in\{0,1\}$ is a random padding-bit. The server then measures in the equatorial basis $\{\ket{+_{\delta_n}},\ket{-_{\delta_n}}\}$ and transmits the outcome to the \gls{que} that decodes the true logical outcome. As the server cannot infer the logical angles, it cannot infer the logical outcome of each measurement, nor the algorithm it runs itself (other than an upper bound on the depth by knowing $K$). The measurements must be done sequentially as $\phi_n$ generally depends on previous measurements, hence the \gls{que} must decode the logical outcome before determining the next logical angle.

As a result, the lifetime of the qubits at the \gls{que} spans the duration of generation, wireless transmission, and measurement. For those at the server, the lifetime includes furthermore the time of entangling them and then the actual computation consisting of a sequence of blocks of measurements and classical communication. 
Notably, the only transmission of qubits is the entanglement distribution, hence the \gls{que} may move out of a quantum cell after having received its qubits and still proceed with the algorithm.

\subsection{Quantum sensing}

As with QKD and BQC, any quantum sensing protocol is prone to errors typically associated with decoherence and dephasing. Similarly to the two former cases, this is primarily due to unwanted interactions between components within the protocol architecture and the surrounding environment. This is typically modelled through the relaxation time $T_1$ and the dephasing time $T_2$, and these system parameters are directly related to the limited sensitivity of the quantum sensing setup, see e.g.~\cite{degen2017quantum}. Since quantum sensors rely on superposition to accumulate phase from a physical signal, any interaction with the environment, such as fluctuations in magnetic fields, degrades the coherence, thus shortening $T_1$ and $T_2$. This environmental noise is particularly prevalent in quantum sensing simply because by definition, sensing relies on the measurement of some external field~\cite{ParLeeHan-22}. Furthermore, decoherence imposes a hard limit on sensing performance. By contrast, in quantum computation, short coherence times may potentially be improved by Quantum Error Correction (QEC) codes.

Quantum sensing imposes stricter latency requirements than other quantum applications. For instance, many current sensors require complete re-initialization before each measurement cycle, aggravating the overall latency. This issue is alleviated in quantum computing, as mid-circuit resets are becoming increasingly possible, reducing the latency caused by state reset~\cite{Dooley_2016}.

\pp{In summary, all three applications share common phases, such as setup, entanglement distribution, local quantum operations, and classical reconciliation. Furthermore, they highlight the need for balancing quantum timing constraints imposed by decoherence with digital timing requirements of real-world applications, clearly showing the interdependence between quantum and classical communication in 1Q systems.}

\section{Architecture and Protocol Functionality}\label{sec:architecture}
\subsection{Architectural Considerations}
The \gls{qbs} associated with a quantum cell will enable, control, and manage access to quantum services and applications, as well as coordinating their employment, where the services rely on classical communications and networking.
The intertwining of quantum and classical communications spans \emph{all} protocol layers in 1Q networks, with interdependencies across the classical and quantum protocol stack~\cite{Cacciapuoti:22}.

The server or the peer for a given application may reside in the same cell, or, more likely, be located remotely in another 1Q cell or a QLAN connected to the QI. There are currently several proposals for the QI protocol stack~\cite{Illiano:22}, but their standardization is far from being complete.

As an access network, 1Q may feature its own quantum protocol stack customized for the 1Q functionalities and, to the extent possible, enabling a non-disruptive combination with the protocol stack of the classical cellular network and reusing the existing infrastructure. 
In the classical segment, resources are organized in the time, frequency, and spatial domains.
For each antenna port, a \gls{prb} is the scheduling unit which defines the smallest contiguous block of time-frequency resource elements that can be allocated to a \gls{cue}.
Moreover, the data plane transports the payload information, while the control plane transports the metadata, i.e., all that is needed to manage and deliver that payload. In the quantum segment, entanglements serve as the primary expendable resource. Although classical and quantum resources occupy different physical substrates, optimal network operation requires a consistent allocation strategy, since the performance of emerging quantum applications critically hinges on classical control protocols.
Noting that the control data of quantum applications is mission-critical, a dedicated network slice can be defined for quantum control signaling within the radio access and the core network. For instance, establishing multipartite entanglement might require a specialized configuration that involves a satellite \gls{qbs}, quantum routers, and switching fabrics.  

It is reasonable to assume that the quantum physical layer in 1Q cells will be different from the one used in the quantum backbone, where the analogy between wireless-fiber optic and FSO-fiber optic is evident. An important distinction is that wireless transmissions rely on broadcasting, whereas FSO transmissions, particularly in the context of entanglement distribution, are of point-to-point nature.
Hence, the \gls{qbs} should also perform the functions of a media converter in the quantum data plane.
In the rest of the section we will employ the terminology of user plane instead of data plane, as it is customary in mobile cellular networks.

\subsection{Access Protocols and Connection Management}

In defining the quantum protocol stack, a key challenge is the variability of decoherence times across entanglement technologies, which imposes strict timing constraints and complicates interoperability with classical access. To highlight these hardware-specific differences, 1Q shall define the fidelity requirements of the quantum applications and standardize the capabilities of the nodes in terms of quantum memories, entanglement purification procedures, and error-correction mechanisms. By buffering, refining, and protecting entangled states, these techniques harmonize diverse fidelities and lifetimes, enabling higher-level protocols to function uniformly. This can be expressed through a quantum device class linked to the fidelity curve of \glspl{que}.

When connecting to the QI, it is expected that the 1Q network will require some sort of protocol translation and tunneling, pertaining to control plane protocols managing the quantum communications and services and to the user plane protocols managing the classical communications and services that support the quantum operations.

\begin{figure}[t!]
 \centering
 \includegraphics[width=0.95\columnwidth]{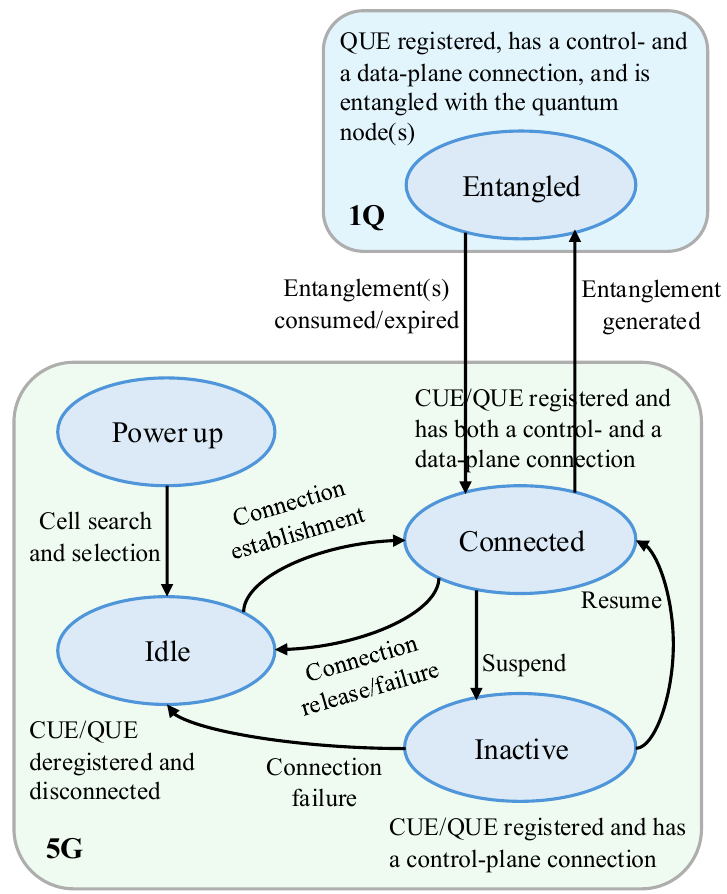}
 \caption{CUE/QUE state diagrams in 5G/1Q access network.}
 \label{fig:5G_1Q_states}
\end{figure}

We now summarize the standard operation of a 5G classical cell, and use it as a blueprint for a quantum cell. 
When powered up, a \gls{cue} performs a cell search to acquire time and frequency synchronization with the nearest \gls{cbs}.
This becomes even more critical, when the quantum physical layer is incorporated, as the temporal information about the quantum resources would naturally serve as a part of their identification, when the network processes multiple qubits concurrently. Following synchronization, the \gls{cue} performs cell selection, camping in the cell within the idle state of Fig.~\ref{fig:5G_1Q_states}.
The \gls{cue} then performs the initial registration procedure, involving a series of exchanges between the \gls{cue} and the \gls{cbs} over a random access channel and dedicated data channels in the uplink and downlink, as well as between the \gls{cbs} and the network core.
During the procedure, the \gls{cue} becomes authenticated. Then encryption keys are generated\footnote{This is another example where a quantum application, i.e., \gls{qkd}, can be used to augment classical networking.}, and the \gls{cue}  is deemed configured, registered, and has a network slice assigned. 
Effectively, the \gls{cue} establishes a connection to the \gls{cbs} in order to register. 
After the registration procedure is completed, the \gls{cue} transitions to the connected state, in which it has both user-- and control-plane connection with the \gls{cbs} and can set up data services.
If the \gls{cue} has no data to send or receive for some time, it transitions back to the idle or inactive state. 
In the inactive state, the user-plane connection to the \gls{cbs} is suspended, but the control-plane connection is retained; this way, the user can faster reestablish the user-plane connection.
Notably, every time the \gls{cue} is in the idle state and requires a data service, it has to transition to the connected state, which is achieved by completing a new connection-establishment procedure.
In summary, mobile cellular networks are connection-oriented.

The connection-oriented approach is a necessity for quantum communications, where classical communications (i.e., signalling) will be needed to setup quantum links and control qubit exchanges~\cite{Illiano:22}.
In 1Q networks, the connection establishment procedure can be naturally extended to incorporate the assignment of a quantum slice and an entanglement generation between the \gls{qbs} and the already (classically) connected \gls{que}.
For this purpose, the \gls{qbs} and the \gls{que} should use a link-layer entanglement generation protocol~\cite{10.1145/3341302.3342070}.
Here, the \gls{qbs} will serve as the node locally generating and distributing quantum resources among \glspl{que}, effectively playing the role of a super-node/quantum-orchestrator, while the \glspl{que} will act as clients~\cite{Mazza:25}, consuming and potentially storing entanglements.
This hierarchical division of the roles, compatible with the organization of the cellular radio-access networks, naturally fits for quantum communications, due to the expected complexity and intricacy of the hardware needed to simultaneously manage multiple quantum states.

Fig.~\ref{fig:5G_1Q_states} shows a potential state diagram of a \gls{que}, featuring an additional state, namely the entangled state. 
Specifically, the \gls{que} can transit from the connected to the entangled state upon successful completion of the link-layer entanglement protocol's session.
This session should involve the following steps: (1) generation of a request (could be initiated both by the \gls{que} and the \gls{qbs}), which should also specify the type of request (i.e., quantum slice), e.g., generate and store or generate and measure, the number of entangled pairs to be created, the maximum tolerable latency of the entanglement generation, application type and priority, minimum acceptable fidelity, etc., (2) a session of a physical layer protocol tasked with producing and distributing the requested entanglement, where the entangled pairs are produced by the \gls{qbs} and shared with the \gls{que}, (3) an acknowledgment by the \gls{que} that the entangled pair has been successfully shared~\cite{10.1145/3341302.3342070}.
Note that steps (1) and (3) are exclusively based on classical signaling.
If the \gls{qbs} is not the end point of the distributed application, the \gls{qbs} has to act as a quantum repeater and perform entanglement swapping with the quantum nodes in the core network or the QI. 
Analogous to 5G, 1Q will have a quantum user plane function (QUPF) to support quantum links among \glspl{que} residing in different 1Q cells, as well as to support quantum links to other communicating parties in the QI.

Although a 1Q cell has a star topology, the \gls{qbs} can create arbitrary artificial topologies comprising quantum links among the \gls{qbs} and/or \glspl{que}, where the entanglements can be bipartite or multipartite.
Moreover, the entangled state can be maintained after a \gls{que} leaves the quantum coverage of the 1Q network, given that the \glspl{que} can store qubits. However, the \gls{que} has to remain in the classical coverage of the 1Q network to be able to complete the service. 

The \gls{qbs} will also have to augment control plane functions with respect to a classical \gls{cbs} to support quantum links.
For instance, radio resource control of such links will have to work in concert with the quantum resource control. 
The current functional architecture of 5G core features a rich set of components, dealing with user registration, authentication, access management, mobility, network slice selection, session management, policy control, etc., offering a platform that could be augmented to support quantum networking. 
Principles that could be used to derive the control plane functions dealing with the quantum networking in the QI were recently outlined in~\cite{CalCal-25}, and could serve as the starting point for the analogous considerations in 1Q systems. 

\glspl{que} are expected to be quasi-static and nomadic rather than mobile, at least in the mid-term, due to hardware complexity and costs.
Once entangled, the parties remain connected in the quantum realm as long as the entanglement is active.
Nevertheless, a \gls{qbs} may be mobile, e.g. satellite, such that network-initiated entanglement handover may be relevant for scenarios in which the entanglement with the serving \gls{qbs} is necessary, as in edge computing.
We may further distinguish between \emph{(i)} soft entanglement handover (make-before-break), realized through entanglement swapping among the \glspl{qbs}; and \emph{(ii)} hard entanglement handover (break-before-make), where the existing entanglement is destroyed, and a new one generated with the new serving \gls{qbs}.
The latter option may be easier for practical implementation, but the former may be a better choice for latency-sensitive applications.
Notably, mobility management in the entangled mode is still required for the classical data.

Finally, the 5G network can act both reactively and proactively when it comes to service execution.
In the traditional, reactive approach, the service is triggered by connection setup when the \gls{cue} wants to send or receive data, paging, network access requests, etc.
This behavior is efficient for conserving energy, minimizing signalling and resource-usage. 
In proactive service execution (that can be combined with predictive capabilities), resources and service functions are pre-allocated, which can reduce latency and maintain quality-of-service.
It is reasonable to assume that 1Q networks would support both paradigms.
In the reactive scenario, \glspl{que} would compete for quantum (and classical) resources with other \glspl{que}, attempting connection-establishment when service-requests arise.
On the other hand, a \gls{qbs} can proactively maintain quantum links (and, if needed, classical links), with guarantees on entanglement-generation rate, fidelity, and link down-time, by assigning users to adequate quantum slices.  

\section{Conclusion}\label{sec:conclusion}

We sketched out a speculative 1Q wireless generation is, which includes quantum information processing into the established cellular architectures. Several key elements of the 1Q network were defined, such as quantum base stations and cells in conjunction with their classical counterparts. Architectural principles and protocol stack requirements were presented to support connection establishment, entanglement distribution, as well as synchronization of quantum and classical network requirements. 
 
 We hypothesized that we are currently at a ``transistor moment'' for quantum technologies and their development will evolve towards larger scale roll-out. 1Q is a system concept that anticipates and, potentially, inspires these development. Yet, its timeline is critically dependent on the pace of evolution in quantum technologies.  Future research includes standardizing protocol specifications for quantum resource allocation, adaptive resource management across heterogeneous devices, scalable network architectures that combine satellite and terrestrial links, and robust quantum error correction tailored for wireless environments. 

\bibliographystyle{IEEEtran}
\bibliography{references}

\end{document}